\documentclass[twocolumn]{IEEEtran}
\usepackage{authblk}
\usepackage{silence}
\usepackage{dblfloatfix} 
\usepackage[font=scriptsize]{caption}

\usepackage[pdftex]{graphicx}
\usepackage{amsmath, amsthm, amssymb}
\usepackage[table,xcdraw]{xcolor}
\usepackage{array}
\usepackage{cite}
\usepackage{amssymb}
\usepackage{color}
\usepackage{amsmath}
\usepackage{multirow}
\usepackage{dirtytalk}
\usepackage{amsmath}
\usepackage{algorithm} 
\usepackage{algpseudocode} 
\usepackage{enumerate}

\begin{document}

\title{Volumetric Lung Nodule Segmentation using Adaptive ROI with Multi-view Residual Learning}
\author[1]{Muhammad Usman}
\author[2]{Byoung-Dai Lee}
\author[1]{Shi Sub Byon}
\author[1]{Sung Hyun Kim}
\author[1]{Byung-il Lee}
\affil[1]{Center for Artificial Intelligence in Medicine and Imaging, HealthHub Co. Ltd., Seoul, 06524, South Korea}
\affil[2]{School of Computer Science and Engineering, Kyonggi University, Suwon, 16227, South Korea}

\affil[ ]{ {usman@healthhub.kr, blee@kyonggi.ac.kr, terry.byon@healthhub.kr, kim.shyun@gmail.com, lbi@healthhub.kr}}                
\setcounter{Maxaffil}{0}
\maketitle

\begin{abstract}
Accurate quantification of pulmonary nodules can assist the early diagnosis of lung cancer, enhancing patient survival possibilities. A number of nodule segmentation techniques, which rely on a radiologist-provided 3-D volume of interest (VOI) input or the use of a constant region of interest (ROI), are proposed; however, these techniques can only investigate the presence of nodule voxels within the given VOI. Such approaches restrain the solutions to freely investigate the nodule presence outside the given VOI and also include the redundant structures (non-nodule) into VOI, which could lead to inaccurate nodule segmentation. In this work, a novel semi-automated approach for 3-D segmentation of nodule in volumetric computerized tomography (CT) lung scans has been proposed. The technique is segregated into two stages. In the first stage, a 2-D ROI containing the nodule is provided as an input to perform a patch-wise exploration along the axial axis using a novel adaptive ROI algorithm. This strategy enables the dynamic selection of the ROI in the surrounding slices to investigate the presence of nodules using a Deep Residual U-Net architecture. This stage provides the initial estimation of the nodule utilized to extract the VOI. In the second stage, the extracted VOI is further explored along the coronal and sagittal axes with two networks . All  the estimated masks are then fed into a consensus module to produce a final volumetric segmentation of the nodule. The proposed approach is rigorously evaluated on LIDC-IDRI dataset, which is the largest publicly available dataset. The results suggest that the proposed approach is more robust and accurate than the previous state-of-the-art  techniques.

\end{abstract}


\section{Introduction}
\label{intro}
Lung cancer, as one of the most severe cancers with high prevalence, which makes it  the leading cause of cancer related death worldwide \cite{torre2015global}. It was forecast to be one of the greatest single cause of mortality among the European population in 2019 \cite{malvezzi2019european}. Early diagnosis of lung cancer is crucial to enable possible life-saving interventions \cite{blandin2017progress} and it relies on accurate quantification of pulmonary nodule; albeit pulmonary nodules can be associated with several diseases, their recurrent diagnosis is lung cancer, in most cases. The continuous monitoring of lung nodule volume is vital to estimate the malignancy and to better forecast, the probability of lung cancer \cite{mozley2012measurement, devaraj2017use}. For calculation of volume, the nodule is first segmented, while the manual segmentation of nodule is a tedious and time-consuming task which also introduces the inter-observer and intra-observer variability \cite{moltz2009advanced}.

\begin{figure}[ht]
\centering
%captionsetup{justification=centering}
\centerline{\includegraphics[width=.5\textwidth]{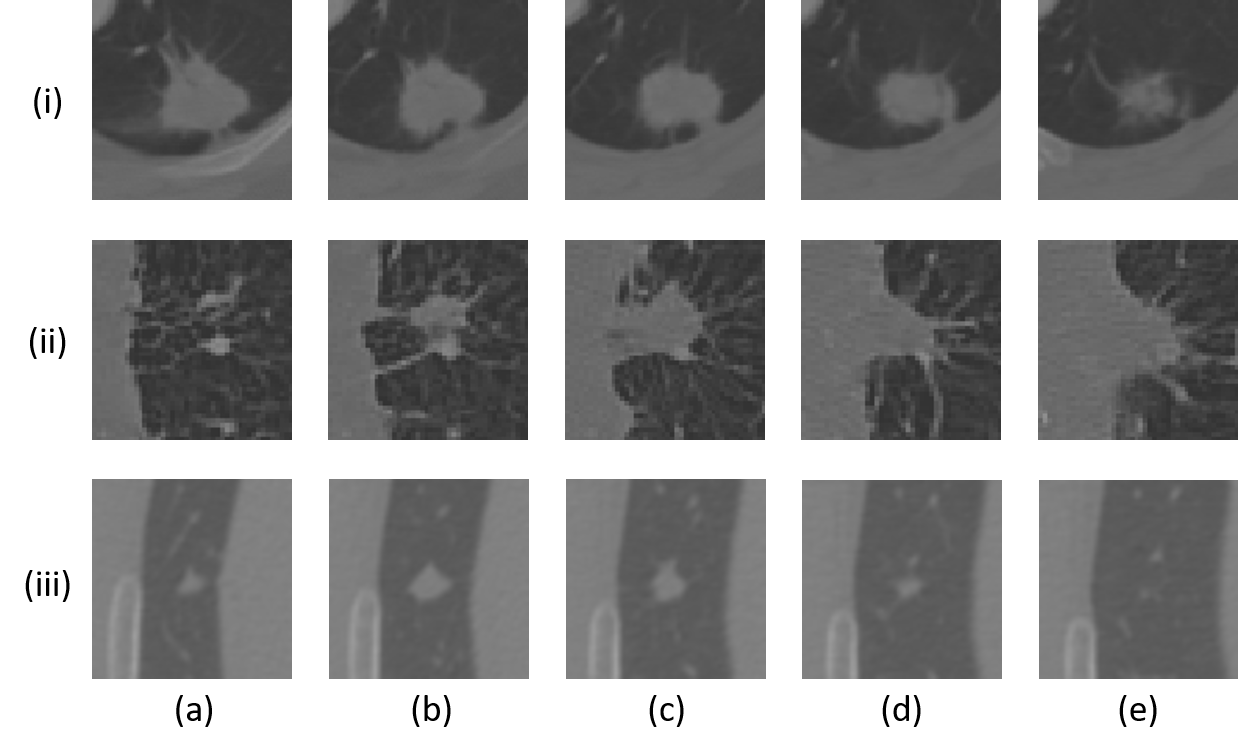}}
\caption{Multiple visual appearances of the pulmonary nodule are shown. The variations within a nodule can be seen from column (a) to (e), and the inter-nodule difference is presented from rows (i) to (iii).}
\label{nodule_type}
\end{figure}

Computer-aided diagnosis (CAD) systems assist in overcoming the challenges faced during manual segmentation of pulmonary nodules and can remarkably enhance the productivity of radiologists. Therefore, several automatic nodule segmentation techniques have been proposed to facilitate radiologists, including advanced deep learning and classical image processing based techniques \cite{wu2019survey}. However, owing to significant shape and contrast variations in lung nodule images, all existing techniques require a 3-D volume of interest (VOI) as an input to precisely estimate the shape of a nodule\cite{wang2017central,shakibapour2019unsupervised,liu2019cascaded}. Specifically, shape variations within a nodule and the visual similarity between a nodule and its surroundings (i.e., non-nodule tissue) act as barriers toward the development of an accurate and robust nodule segmentation solution. Figure \ref{nodule_type} illustrates intra-nodule and inter-nodule variations, wherein the diversity between the shapes of different nodules and variations within a single nodule is observable.

Challenges associated with providing 3-D VOI as input can result in severe drawbacks for exiting solutions, such as limited accuracy and tedious utilization. Since the radiologist provides the initial VOI, the subject involvement is enhanced, which subsequently increases uncertainty in the results due to inter- and intra-human judgement. On the other hand, a radiologist investigates the presence of a pulmonary nodule using 2-D views (i.e., axial, coronal, and sagittal), so providing a 3-D VOI becomes additionally laborious task.

In this work, we propose a novel approach to perform volumetric segmentation of pulmonary nodules by taking only the 2-D region of interest (ROI) input from the radiologist. The solution first explores the presence of a nodule within the provided ROI by employing a Deep Residual U-Net and then extends the search into surrounding slices (in both directions). To investigate possible penetration of the nodule within adjacent slices, we introduce the novel concept of adaptive ROI (A-ROI) that allows the solution to dynamically change the position and size of ROI while searching into other slices. The application of this A-ROI algorithm along the axial plane provides an initial estimation of the nodule shape, which is then used  to extract a 3-D VOI from the scan automatically. This VOI is further utilized to create coronal and sagittal views of the nodule, and slices of both are investigated with two Deep Residual U-Nets. Finally, three estimated segmentation masks (i.e., for the axial, coronal, and sagittal views) are fed into a consensus module to build a final segmentation mask. To validate the performance of the proposed method, an extensive set of experiments has been conducted on LIDC-IDRI dataset \cite{armato2011lung}, which is the largest publicly available dataset. The results suggest that the approach is robust and significantly improves the performance in terms of dice score as compared to the previous state of the art techniques.

The rest of the paper is organized as follows. In section \ref{Re}, we present background and related work. In section \ref{pro}, the proposed method and its all the steps are described in detail. Section \ref{ES} explains the experimental procedure and in section \ref{RD} obtained results are discussed. Finally, we conclude in section \ref{co}.

% \begin{figure*}[ht!]
% \centering
% %captionsetup{justification=centering}
% \centerline{\includegraphics[width=.9\textwidth]{proposed_method.png}}
% \caption{The stages of proposed method have been shown, at stage I, manual ROI along axial axis is provided by user and deep Residual U-Net along with adaptive ROI algorithm is employed to volume of interest (VOI). After VOI extraction at stage II, patch-wise segmentation of nodule is performed along coronal and sagittal axis. Eventually, the consensus module is employed on all the estimated segmentation masks to obtain the final 3-D segmentation mask nodule.}
% \label{propose_fig}
% \end{figure*}

\section{Related Work}
\label{Re}

An accurate assessment of the lung nodule is required to investigate its malignancy and possibility of lung cancer. Because of the extraordinary importance of nodule segmentation, several efforts have been made to develop a highly accurate and robust nodule segmentation system that can assist radiologists. These efforts may be distinguished into two categories: deep learning-based techniques and classical image processing-based techniques\cite{wu2019survey}. In this section, we incorporate a brief review of recently proposed techniques from each category.

Jamshid et al.\cite{dehmeshki2008segmentation} proposed an algorithm that utilized two-region growing techniques (i.e., fuzzy connectivity and contrast-based region growing) to perform nodule segmentation. The region growing is operated within a volumetric mask that is created by first applying a local adaptive segmentation algorithm to identify the foreground and background regions within a specified window size. The algorithm performed well for a separated nodule but failed to segment the attached nodules. Stefano et al.\cite{diciotti20083} proposed a user interactive algorithm that adopts geodesic influence zones in a multi-threshold image representation to allow the achievement of fusion–segregation criteria based on both gray-level similarity and objects shape. The same author extended this work\cite{diciotti2011automated} by removing the user interaction component and performing corrections according to 3-D local shape analysis. The correction procedure refined an initial nodule segmentation to split possible vessels  from the nodule segmentation itself. 

Threshold and morphological techniques were adopted in Elmar et al.\cite{rendon2016automatic} to eliminate the background and other surrounding information from the provided ROI. Then, a support vector machine (SVM) was employed to classify each pixel in the detected space. Similarly, Wang et al.\cite{wang2017adaptive} tried to segment solitary pulmonary nodules in digital radiography (DR) images by incorporating a sequential filter to construct new representations of the weight and probability matrices. However, this method is limited to DR images, which constrains the application for CT images. Additionally, Julip et al.\cite{jung2018ground} segmented ground-glass nodules (GGN) in chest CT images using an asymmetric multi-phase deformable model. However, this technique lacks the robustness to address segmentation requirements for other nodule types.

Shakibapour et al.\cite{shakibapour2019unsupervised} employed the notion of optimally clustering a set of feature vectors comprising intensity and shape-related features in a given feature data space extracted from a predicted nodule. The size is obtained by measuring the volume of a segmented nodule via an ellipsoid approximation using the equivalent diameters of the segmented regions in a 2.5-D representation; thus, the uncertainty persists into the final results. Shakir et al.\cite{shakir20183} proposed the voxel intensity-based segmentation scheme that incorporates mean intensity-based thresholding in the Geodesic Active contour model in level sets. This work was validated on limited set of scans, so the robustness of the proposed technique is dubious.

%In \cite{shakibapour2019unsupervised}, Shakibapour et al. employed the idea to optimally cluster a set of feature vectors composed by the intensity and shape-related features in a given feature data space extracted from a predicted nodule. The nodule size is obtained by measuring the volume of a segmented nodule via ellipsoid approximation using the equivalent diameters of the segmented regions in a 2.5D representation, which leaves the uncertainty into the final results. In \cite{shakir20183}, Shakir et al proposed the voxel intensity-based segmentation scheme which incorporates mean intensity-based thresholding in the Geodesic Active contour model in level sets. The work was validated on only 72 scans which dubious the robustness of the proposed technique.

\begin{figure*}[ht]
\centering
%captionsetup{justification=centering}
\centerline{\includegraphics[width=.9\textwidth]{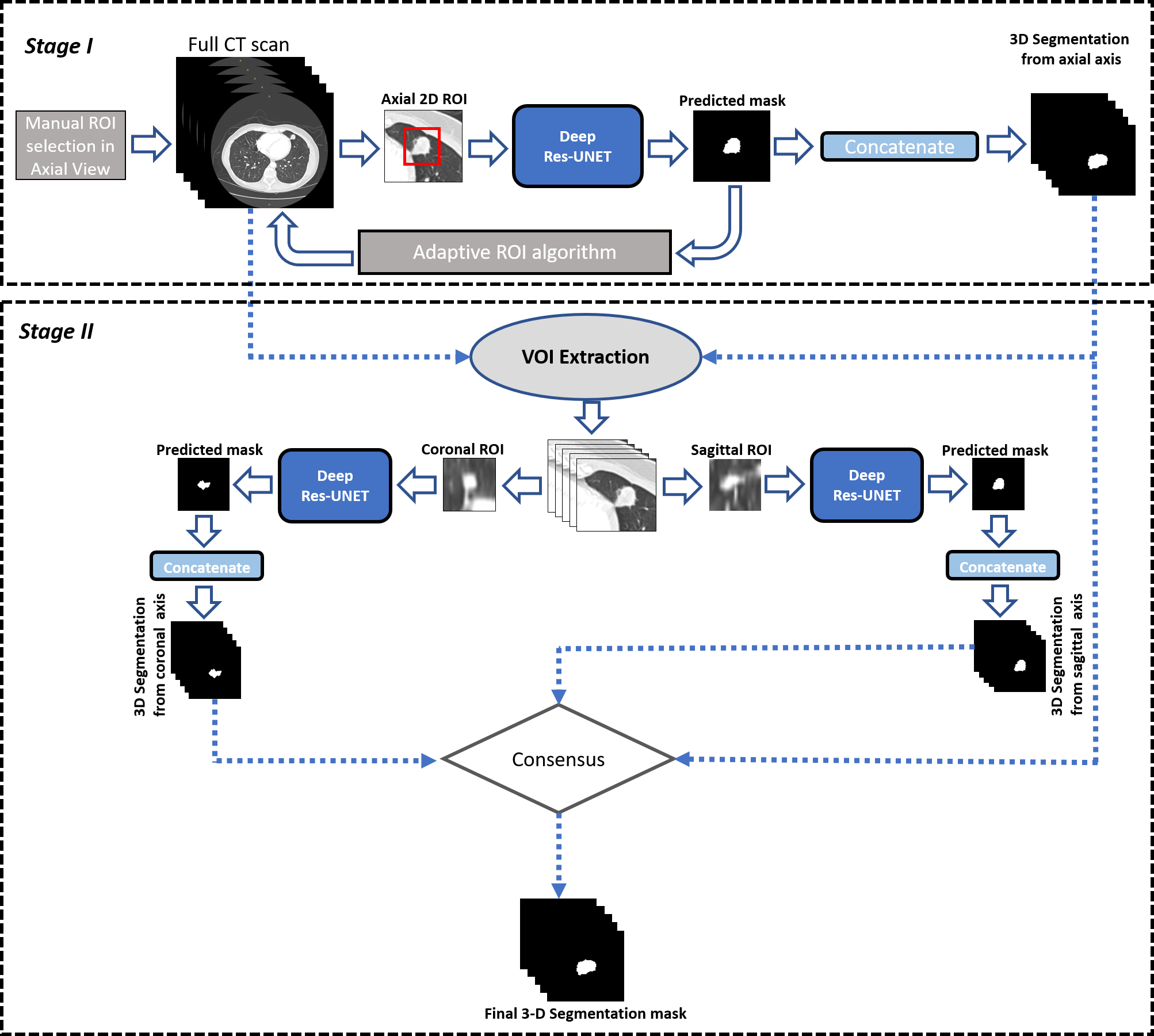}}
\caption{An illustration of the stages of the proposed method. At stage I, a manual ROI along the axial axis is provided by the user, and a Deep Residual U-Net along with the adaptive ROI algorithm is employed to extract the volume of interest (VOI). After getting the VOI during stage II, a patch-wise segmentation of the nodule is performed along the coronal and sagittal axes. Eventually, a consensus module is employed on all estimated segmentation masks to obtain the final 3-D nodule segmentation mask.}
\label{propose_fig}
\end{figure*}

Despite extensive research, classical image processing techniques are unable to provide sufficiently robust and accurate volumetric nodule segmentation. On the other hand, recent advancements in deep learning \cite{latif2018phonocardiographic,latif2018mobile,usman2017using,latif2018cross,latif2018automating,usman2019motion} have revolutionized image segmentation-related applications\cite{litjens2017survey,guo2018review,usman2019volumetric}, including lung nodule segmentation tasks\cite{rocha2019comparison}. Especially the introduction of the U-Net architecture\cite{ronneberger2015u} for segmentation in medical images remarkably enhanced the performance for these tasks. Therefore, many efforts have been made recently to exploit deep learning for lung nodule segmentation, such as Wang et al.,\cite{wang2017central} who developed the Central Focused Convolutional Neural Network (CF-CNN) that takes a volumetric patch around a selected voxel as the input and returns the segmentation of nodule. The same authors in\cite{wang2017multi} presented a multi-view CNN for lung nodule segmentation, which takes the axial, coronal and sagittal view around a given voxel of a nodule as input and provides nodule segmentation. The method of patch (axial, coronal, and sagittal) extraction around the given nodule is kept fixed for all nodules which may lead to inaccurate segmentation if the nodule is larger than the extracted patch size.

Guofeng et al.\cite{tong2018improved} improved the performance of U-Net for nodule segmentation by including skipped connections within the encoder and decoder paths. Enhanced performance of this proposed architecture is reported, but the model is only designed for 2-D segmentation. Similarly, Amorim et al.\cite{amorim2019lung} modified the U-Net architecture to investigate the presence of nodules through a patch-wise approach. They manually extracted a VOI containing the nodule and performed 2-D segmentation along each axis, so a final segmentation is calculated by summing the three predicted masks. The VOI selection criterion is the disadvantage of this method, as it is static and is fixed as a 128 x 128 x 12 window around the given voxel. 
Hancock et al.\cite{hancock2019lung} presented an extension of the vanilla level set image segmentation method in which instead of being manually designed, the velocity function is learned from data via machine learning regression methods. They employed their segmentation scheme for lung nodule segmentation and reported slightly improved performance. A residual block based dual-path network in Liu et al.\cite{liu2019cascaded} extracts local features and rich contextual information of lung nodules, which resulted in a significant performance enhancement. However, they also used a fixed VOI that constrains  the search of the nodule and subsequently decreases the performance in 3D segmentation.

In this work, we eliminate the downsides of the fixed VOI by introducing the adaptive 2-D ROI selection algorithm, which greatly assists to exploit the power of deep learning. We also utilize the Deep Residual U-Net\cite{zhang2018road}, which has shown excellent performance with other segmentation tasks but has never been incorporated into nodule segmentation. We further investigate the automatically extracted VOI along the coronal and sagittal axes to determine an accurate segmentation of a lung nodule.

\section{Proposed Method}

\label{pro}
Our method consists of two stages, as described in Figure \ref{propose_fig}. In the first stage, we estimate the nodule 3-D shape along the axial axis to extract the VOI. In the second stage, we utilize the extracted VOI to further perform a patch-wise investigation along the sagittal and coronal axes. Finally, we use the consensus module to calculate the 3D segmentation of the nodule. Details of each stage are described below.

\subsection{Stage I}

The 2-D ROI containing a nodule is manually provided by the radiologist and may be selected from any portion of the nodule (i.e., from any slice). This ROI is then processed by the Deep Residual U-Net architecture to obtain a 2-D segmentation of the nodule, which is next forwarded into the A-ROI  algorithm; the algorithm utilizes the position of nodule within the current ROI to determine the size and position of the ROI for the subsequent slice. Finally, the volumetric segmentation mask of the nodule is built by concatenating all the 2-D segmentation masks.

Descriptions of the A-ROI algorithm and Deep Residual U-Net architecture are provided in the following sections.

\subsubsection{Adaptive ROI Algorithm}
\label{aroi-section}

We propose a novel algorithm for the dynamic selection of the ROI in each slice to search for the presence of a nodule. The objectives of this algorithm include:

\begin{enumerate}[i.]
\item To maintain the predicted nodule mask and ROI concentric by adjusting the position of the ROI.
\item To maintain the ratio of the areas of the nodule and ROI below a threshold level (RT), by adjusting the size of the ROI.
\end{enumerate}

\begin{figure}[h]
\centering
%captionsetup{justification=centering}
\centerline{\includegraphics[width=.25\textwidth]{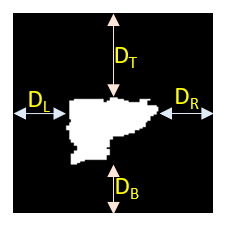}}
\caption{The margins on the four sides of the predicted mask are highlighted as $D_L, D_R, D_T,$ and $D_B$, for the left, right, top, and bottom margins of the predicted nodule segmentation, respectively.}
\label{DROI_mask}
\end{figure}

\begin{figure*}[h]
\centering
%captionsetup{justification=centering}
\centerline{\includegraphics[width=.9\textwidth]{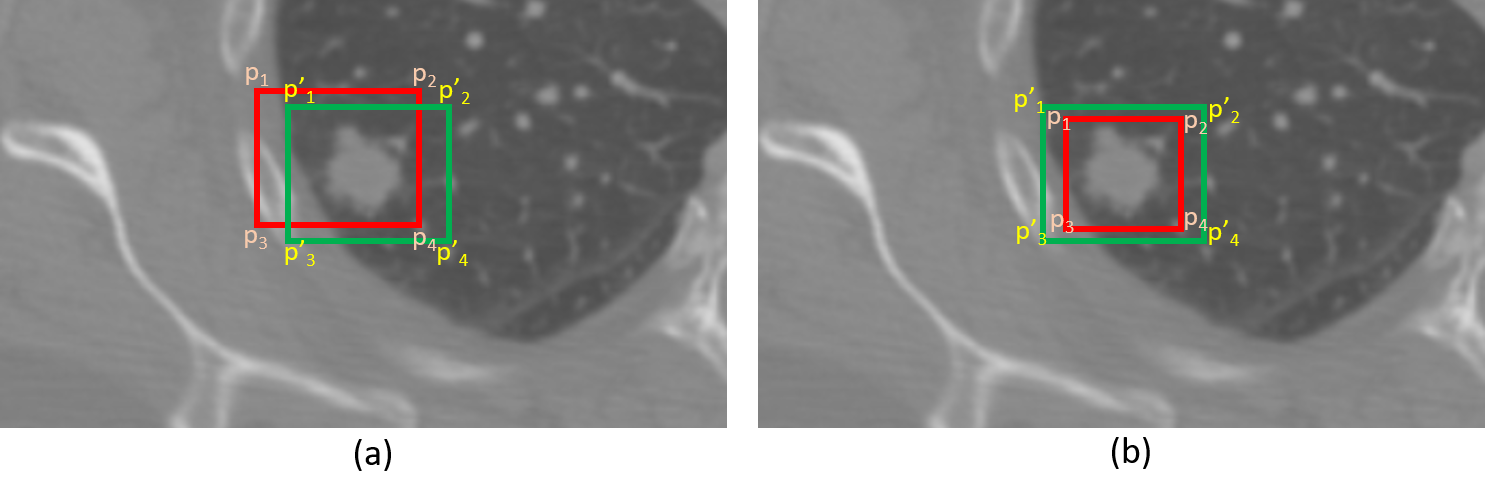}}
\caption{The change in position and shape of the ROI after employing our A-ROI algorithm, is presented. The red square indicates the position of the ROI for the current slice, and the green square represents the ROI for the subsequent slice. Points $p_1(x_1,y_1)$, $p_2(x_2,y_1)$, $p_3(x_1,y_2)$, and $p_4(x_2,y_2)$ belong to ROI of current slice and points $p'_1(x_1,y_1)$, $p'_2(x_2,y_1)$, $p'_3(x_1,y_2)$, and $p'_4(x_2,y_2)$ points belong to the subsequent slice.}
\label{DROI_show}
\end{figure*}

Algorithm \ref{algo_1} describes the steps of the nodule penetration search into adjacent slices. First, nodule segmentation is performed on the manual ROI provided by the user. After obtaining a predicted segmentation of a nodule, the ROI position and size are adjusted by considering the margins identified in the predicted mask.

In Algorithm \ref{algo_1}, $A_{N_i}$, $ A_{ROI_i}$, and $A_{ROI_{i+1}}$ indicate the area of a nodule in the current ($i_{th}$) slice, the area of the ROI in the current ($i_{th}$) slice, and the suggested area of ROI in next ($(i\pm1)_{th}$) slice, respectively. $D_L$, $D_R$, $D_T$, and $D_B$ are the left, right, top and bottom margins of the predicted mask, respectively, as shown in Figure \ref{DROI_mask}, while $\Delta D_X$ and $\Delta D_Y$ provide the differentials in the margins along the x and y axes, respectively. $X_1, X_2$ and $Y_1, Y_2$ are the beginning and ending coordinate points along the $x$ and $y$ axes of the current ROI, and $X'_1, X'_2$ and $Y'_1, Y'_2$ are the beginning and ending coordinate points along the $x$ and $y$ axes of the updated ROI in the next slice ($S_{i\pm1}$), respectively, as shown in Figure \ref{DROI_show}.

\begin{algorithm} [!b]
% \scriptsize
	\caption{Dynamic ROI Selection}
	\begin{algorithmic}[1]
	\State Initial ROI provided of user
	\State Nodule segmentation in current slice ($S_i$) with given ROI using Deep residual U-Net
	
		\While {$A_{N_i} > 0$}
		\State $\Delta D_X = D_L - D_R$
		\State $\Delta D_Y = D_T - D_B$
		\State $X'_1 = X_1 + \Delta D_X$
		\State $X'_2 = X_2 + \Delta D_X$
		\State $Y'_1 = Y_1 + \Delta D_Y$
		\State $Y'_2 = Y_2 + \Delta D_Y$
	
		\If {$\frac{A_{N_i}}{A_{ROI_i}}> R_T$}
		\State $A_{ROI_{i\pm1}} = \frac{A_{N_i}}{R_T}$
		\State $\Delta A = A_{ROI_{i\pm1}} - A_{ROI_i}$
		\State $D_s = ceil(\sqrt{\Delta A}/2)$
		\State $X'_1 \leftarrow X'_1 - D_s$
		\State $X'_2 \leftarrow X'_2 + D_s$
		\State $Y'_1 \leftarrow Y'_1 - D_s$
		\State $Y'_2 \leftarrow Y'_2 + D_s$
		\EndIf
		\State Moving to next slice: $S_i \leftarrow S_{i\pm1}$
		\State Nodule segmentation in next slice with updated ROI using Deep Residual U-Net
		\EndWhile
	\end{algorithmic}
	\label{algo_1}
\end{algorithm} 

After determining the position of the next ROI, Algorithm \ref{algo_1} determines the optimal size of the ROI. The shape of the ROI remains square, so that each side of the calculated ROI has the same length. The algorithm retains the ratio between $A_N$ and $A_{ROI}$, which is less than the constant value of the selected ratio threshold ($R_T$). The value of $R_{T}$ is crucial, as it determines the size of the ROI to be selected for the next slice, which is directly linked to the maximum possible movement of the nodule within two adjacent slices. The size of the ROI should be adequate to address the maximum possible displacement of the nodule that is directly proportional to the slice thickness ($ST$). Therefore,

\begin{equation}
    R_T \propto \frac{1}{ST}
\end{equation}

The optimum value of $R_T$ is determined experimentally and is discussed in the Results Section. Eventually, the A-ROI algorithm tends to retain the condition,

\begin{equation}
    R_T \leq \frac{A_N}{A_{ROI}}
\end{equation}
Here, the minimum possible value of $A_{ROI}$ can be equal to $A_N$, so $R_T \in (0,1) $. 

If the ratio of $A_{N_i}$ and $A_{ROI_i}$ becomes greater than $R_T$, then the algorithm calculates the differential ($\Delta A$) between the current area of the ROI ($A_{ROI_i}$) and the required area of the ROI for next slice ($A_{ROI_{i+1}}$). Then, $\Delta A$ is utilized for updating the coordinates of the ROI to obtain the required size.

In Figure \ref{DROI_show}, the ROI within the current slice and the estimated ROI (calculated using the A-ROI algorithm) for the next slice are shown in red and green, respectively. The change in the position of the ROI is presented in Figure \ref{DROI_show}(a), and in \ref{DROI_show}(b), the change of the size of the ROI is depicted.

\begin{figure*}[t]
\centering
%captionsetup{justification=centering}
\centerline{\includegraphics[width=1\textwidth]{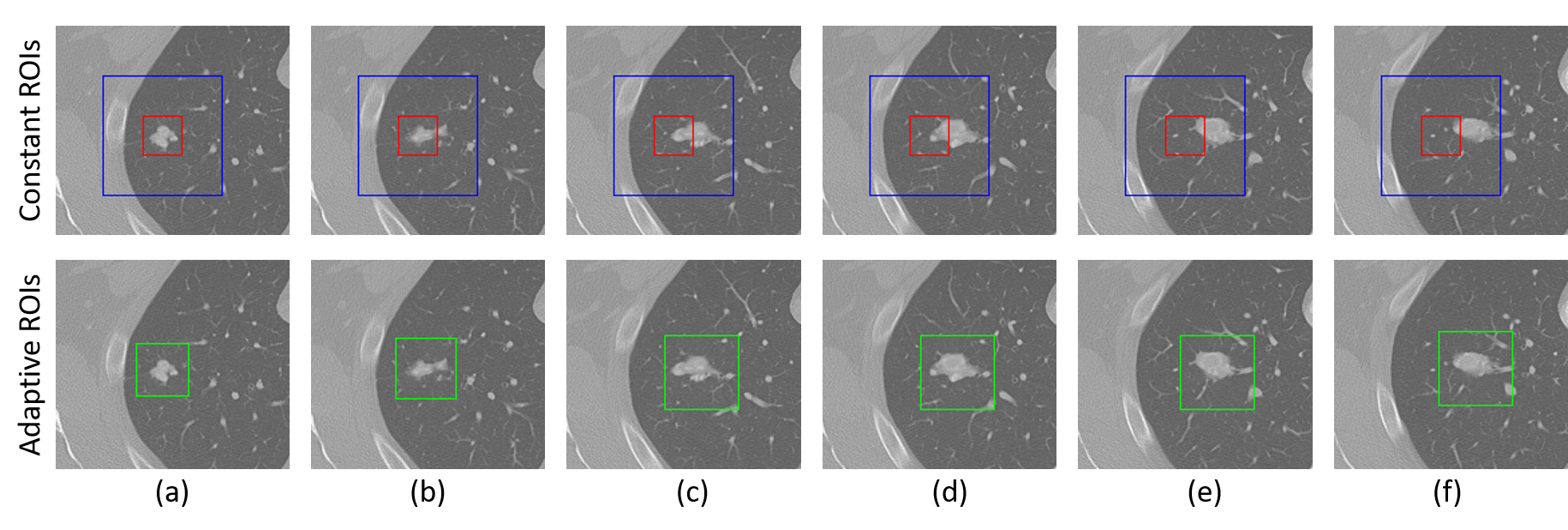}}
\caption{The constant and adaptive region of interests (ROIs) have been shown in a sequence of slices in which nodule is present. Blue and red boxes represent the constant ROIs, while green boxes depict the adaptive ROI.}
\label{ROIs}
\end{figure*}

The entire effect of the A-ROI algorithm is observed in Figure \ref{ROIs}, where two constant ROIs and an adaptive ROI are shown. The segmentation of the nodule starts from column (a) with the manual ROI and ends at column (f). The conventional ROIs (i.e., in red and blue) are the same in each slice while the adaptive ROIs, presented in green, feature different positions and sizes in each slice. In a conventional approach when the ROI remains close to the nodule in the initial slice (as shown in red), it fails to cover the nodule area present in the other slices. Then, a constant and larger ROI (shown in blue ) is required to incorporate the complete area of a nodule spanning all slices. This approach also adds redundant information into the ROIs, which impacts the performance of the segmentation model. On the other hand, our adaptive ROI strategy not only enables the user to select the ROI without the need to look into other slices, but also optimally chooses the ROIs for the remainder of the slices to facilitate improved performance of the segmentation model.

%The selection of ROI while searching the nodule voxels is of prime importance as it can directly effect the nodule quantification. The the thickness of slices in any given CT scan may vary, and most of the times the position of nodule also varies from one slice to another. Therefore, to accurately the segment the nodule in the adjacent slices the position and shape of ROI should be change. In this we propose a novel strategy to estimate the position and area of ROI for any given slice $I_n$ on the basis of predicted segmentation of nodule in previous slice $I_{n-1}$ 

\begin{figure*}[!b]
\centering
%captionsetup{justification=centering}
\centerline{\includegraphics[width=1\textwidth]{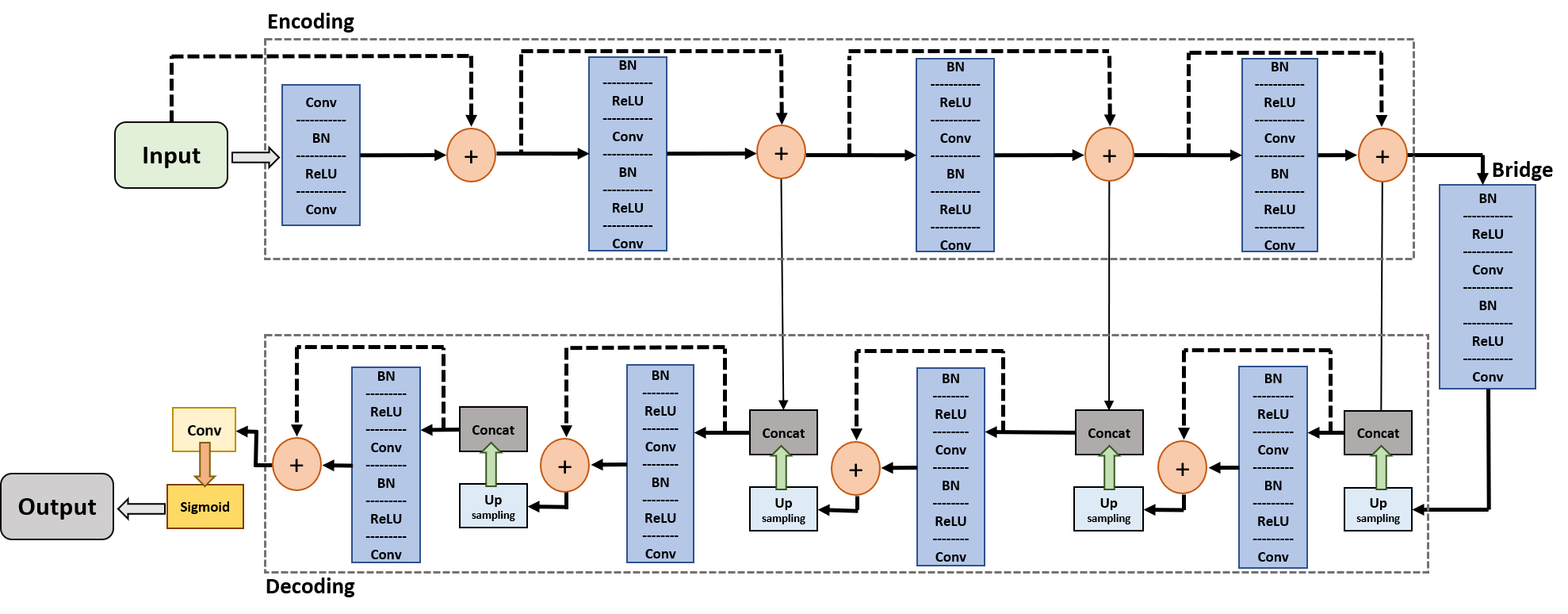}}
\caption{The architecture of Deep Residual U-Net, which is employed along the axial axis with the A-ROI algorithm to perform the lung nodule segmentation.}
\label{deep_Res_UNET}
\end{figure*}

\subsubsection*{Deep Residual U-Net Architecture}
For nodule segmentation in the provided and calculated ROIs, we employ the Deep Residual U-Net architecture\cite{zhang2018road}. The network utilizes the residual learning of the U-Net architecture, which is state of the art in segmentation models. The involvement of residual units eases the training process, and the skip connections allow the flow of information without degradation from low levels to high levels of the network. This combination enables the network to learn the patterns with fewer parameters, which is significant for our application, and increases the performance on conventional U-Net architectures.

We leverage a 9-level architecture of the Deep Residual U-Net , as shown in Figure \ref{deep_Res_UNET}. The network consists of three sections for encoding, a bridge, and decoding. The encoding section contracts the input image into a compact representation. The decoding section recovers the required information (i.e., the semantic segmentation) into a pixel-wise representation. The bridge connects these two sections. All sections of the network are built with residual units consisting of two 3 $\times$ 3 convolution blocks that include a batch normalization layer, a ReLU activation layer, and a convolutional layer. The identity mapping connects the input and output of the unit.

The encoding path has four residual units. In each unit, instead of using a pooling operation to down-sample the feature map size, a stride of 2 is applied to the first convolution block to reduce the feature map by half. Correspondingly, the decoding path also comprises four residual units. Preceding each unit, an up-sampling of the feature maps occurs from the lower level, and a concatenation of the feature maps is applied from the corresponding encoding path. After the final level of the decoding path, a 1$\times$1 convolution and sigmoid activation layer is used to project the multi-channel feature maps into the desired segmentation. The parameters and output sizes of each step are listed in Table \ref{unet_detail}.

\begin{table}[h]
\caption{The network structure of the Deep Residual U-Net that performs patch-wise segmentation along the axial axis.}
\centering
\scalebox{1}{
\scriptsize
\begin{tabular}{lcllcc}
\hline
                & \textbf{Unit level} & \multicolumn{1}{c}{\textbf{Conv Layer}}                   & \multicolumn{1}{c}{\textbf{Filter}}                             & \textbf{Stride}                               & \textbf{Output size}                                                    \\ \hline
\textbf{Input}  &                     &                                                           &                                                                 &                                               & 128 $\times\:$128 $\times\:$1                                                          \\ \hline
                & Level 1             & \begin{tabular}[c]{@{}l@{}}Conv 1\\ Conv 2\end{tabular}   & \begin{tabular}[c]{@{}l@{}}3 $\times\:$3/64\\ 3 $\times\:$3/64\end{tabular}     & \begin{tabular}[c]{@{}c@{}}1\\ 1\end{tabular} & \begin{tabular}[c]{@{}c@{}}128 $\times\:$128$\times\:$64\\ 128 $\times\:$128 $\times\:$64\end{tabular} \\ \hline
                & Level 2             & \begin{tabular}[c]{@{}l@{}}Conv 3\\ Conv 4\end{tabular}   & \begin{tabular}[c]{@{}l@{}}3 $\times\:$3/128\\ 3 $\times\:$3/128\end{tabular}   & \begin{tabular}[c]{@{}c@{}}2\\ 1\end{tabular} & \begin{tabular}[c]{@{}c@{}}64 $\times\:$64 $\times\:$128\\ 64 $\times\:$64 $\times\:$128\end{tabular}   \\ \hline
                & Level 3             & \begin{tabular}[c]{@{}l@{}}Conv 5\\ Conv 6\end{tabular}   & \begin{tabular}[c]{@{}l@{}}3 $\times\:$3/256\\ 3 $\times\:$3/256\end{tabular}   & \begin{tabular}[c]{@{}c@{}}2\\ 1\end{tabular} & \begin{tabular}[c]{@{}c@{}}32 $\times\:$32 $\times\:$256\\ 32 $\times\:$32 $\times\:$256\end{tabular}   \\ \hline
                & Level 4             & \begin{tabular}[c]{@{}l@{}}Conv 7\\ Conv 8\end{tabular}   & \begin{tabular}[c]{@{}l@{}}3 $\times\:$3/512\\ 3 $\times\:$3/512\end{tabular}   & \begin{tabular}[c]{@{}c@{}}2\\ 1\end{tabular} & \begin{tabular}[c]{@{}c@{}}16 $\times\:$16 $\times\:$512\\ 16 $\times\:$16 $\times\:$512\end{tabular}   \\ \hline
                & Level 5             & \begin{tabular}[c]{@{}l@{}}Conv 9\\ Conv 10\end{tabular}  & \begin{tabular}[c]{@{}l@{}}3 $\times\:$3/1024\\ 3 $\times\:$3/1024\end{tabular} & \begin{tabular}[c]{@{}c@{}}2\\ 1\end{tabular} & \begin{tabular}[c]{@{}c@{}}8 $\times\:$8 $\times\:$1024\\ 8 $\times\:$8 $\times\:$1024\end{tabular}     \\ \hline
                & Level 6             & \begin{tabular}[c]{@{}l@{}}Conv 11\\ Conv 12\end{tabular} & \begin{tabular}[c]{@{}l@{}}3 $\times\:$3/512\\ 3 $\times\:$3/512\end{tabular}   & \begin{tabular}[c]{@{}c@{}}1\\ 1\end{tabular} & \begin{tabular}[c]{@{}c@{}}16 $\times\:$16 $\times\:$512\\ 16 $\times\:$16 $\times\:$512\end{tabular}   \\ \hline
                & Level 7             & \begin{tabular}[c]{@{}l@{}}Conv 13\\ Conv 14\end{tabular} & \begin{tabular}[c]{@{}l@{}}3 $\times\:$3/256\\ 3 $\times\:$3/256\end{tabular}   & \begin{tabular}[c]{@{}c@{}}1\\ 1\end{tabular} & \begin{tabular}[c]{@{}c@{}}32 $\times\:$32 $\times\:$256\\ 32 $\times\:$32 $\times\:$256\end{tabular}   \\ \hline
                & Level 8             & \begin{tabular}[c]{@{}l@{}}Conv 15\\ Conv 16\end{tabular} & \begin{tabular}[c]{@{}l@{}}3 $\times\:$3/128\\ 3 $\times\:$3/128\end{tabular}   & \begin{tabular}[c]{@{}c@{}}1\\ 1\end{tabular} & \begin{tabular}[c]{@{}c@{}}64 $\times\:$64 $\times\:$128\\ 64 $\times\:$64 $\times\:$128\end{tabular}   \\ \hline
                & Level 9             & \begin{tabular}[c]{@{}l@{}}Conv 17\\ Conv 18\end{tabular} & \begin{tabular}[c]{@{}l@{}}3 $\times\:$3/64\\ 3 $\times\:$3/64\end{tabular}     & \begin{tabular}[c]{@{}c@{}}1\\ 1\end{tabular} & \begin{tabular}[c]{@{}c@{}}128 $\times\:$128 $\times\:$64\\ 128 $\times\:$128 $\times\:$64\end{tabular} \\ \hline
\textbf{Output} &                     & Conv 19                                                   & 1 $\times\:$1                                                           & 1                                             & 128 $\times\:$128 $\times\:$1                                                          \\ \hline
\end{tabular}
}
\label{unet_detail}
\end{table}

\subsubsection{Loss function}

Given a set of training images and the corresponding ground truth segmentations, ${I_i, s_i}$, we estimate the parameter W of the network, such that it produces accurate and robust nodule segmentation masks. This optimal value is achieved through minimizing the loss between the segmentations generated by $Net(I_i,W)$ and the ground truth $s_i$. We use the dice similarity coefficient (DSC)\cite{zou2004statistical} as the loss function,

\begin{equation}
\textit{L(W)} = \frac{1}{N}\sum_{i=1}^{N} \left[ 1 - \frac{2*Net(I_i;W)\cap s_i}{Net(I_i;W) + s_i}\right]
\label{eq_2}
\end{equation}
\newline
Where $N$ is the number of training samples. We use stochastic gradient descent (SGD) to train our network. 

\subsection{Stage II}
The second stage of our approach is further designated into two phases. The first employs patch-wise nodule segmentation along the coronal and sagittal axes, and the second reconstructs the final 3-D segmentation mask of the nodule from all estimated nodule segmentations. These phases are detailed in the following.

%\subsubsection{VOI Extraction}
%At first phase the presence of nodule is investigated using patch-wise strategy along coronal and sagittal planeand we obtain initial 3D segmentation of nodule. On the basis of first estimated nodule segmentation, Volume of interest (VOI) with certain ratio (RT) is extracted from full scan, which is used to create the view along coronal and sagittal planes of nodule. 

\subsubsection{Multi-view Investigation}
The VOI extracted from stage I is utilized to perform patch-wise nodule segmentation along the coronal and sagittal axes independently with two networks. By considering the fact that slice thickness is often greater than the voxels spacing in the x-y plane, we resize the coronal and sagittal patches to $128\times64$. The segmentation is performed using a similar Deep Residual U-Net architecture as that used during stage I. However, owing to the smaller size of the images, the number of levels is reduced to seven (i.e., three encoding, one bridge, and three decoding units), which decreases the number of parameters in the network; this is shown in Table \ref{unet_detail_1}. The output of each model is resized to the original size of the ROI and then concatenated in the same manner as it was performed during stage I to reconstruct the 3D segmentation of a nodule.

\begin{table}[h]
\caption{The network structure of the Deep Residual U-Net to perform patch-wise segmentation along the coronal and sagittal axes.}
\centering
\scalebox{1}{
\scriptsize
\begin{tabular}{lcllcc}
\hline
                & \textbf{Unit level} & \multicolumn{1}{c}{\textbf{Conv Layer}}                   & \multicolumn{1}{c}{\textbf{Filter}}                           & \textbf{Stride}                               & \textbf{Output size}                                                  \\ \hline
\textbf{Input}  &                     &                                                           &                                                               &                                               & 128 $\times\:$64 $\times\:$1                                                          \\ \hline
                & Level 1             & \begin{tabular}[c]{@{}l@{}}Conv 1\\ Conv 2\end{tabular}   & \begin{tabular}[c]{@{}l@{}}3 $\times\:$3/64\\ 3 $\times\:$3/64\end{tabular}   & \begin{tabular}[c]{@{}c@{}}1\\ 1\end{tabular} & \begin{tabular}[c]{@{}c@{}}128 $\times\:$64 $\times\:$64\\ 128 $\times\:$64 $\times\:$64\end{tabular} \\ \hline
                & Level 2             & \begin{tabular}[c]{@{}l@{}}Conv 3\\ Conv 4\end{tabular}   & \begin{tabular}[c]{@{}l@{}}3 $\times\:$3/128\\ 3 $\times\:$3/128\end{tabular} & \begin{tabular}[c]{@{}c@{}}2\\ 1\end{tabular} & \begin{tabular}[c]{@{}c@{}}64 $\times\:$32 $\times\:$128\\ 64 $\times\:$32 $\times\:$128\end{tabular} \\ \hline
                & Level 3             & \begin{tabular}[c]{@{}l@{}}Conv 5\\ Conv 6\end{tabular}   & \begin{tabular}[c]{@{}l@{}}3 $\times\:$3/256\\ 3 $\times\:$3/256\end{tabular} & \begin{tabular}[c]{@{}c@{}}2\\ 1\end{tabular} & \begin{tabular}[c]{@{}c@{}}32 $\times\:$16 $\times\:$256\\ 32 $\times\:$16 $\times\:$256\end{tabular} \\ \hline
                & Level 4             & \begin{tabular}[c]{@{}l@{}}Conv 7\\ Conv 8\end{tabular}   & \begin{tabular}[c]{@{}l@{}}3 $\times\:$3/512\\ 3 $\times\:$3/512\end{tabular} & \begin{tabular}[c]{@{}c@{}}2\\ 1\end{tabular} & \begin{tabular}[c]{@{}c@{}}16 $\times\:$8 $\times\:$512\\ 16 $\times\:$8 $\times\:$512\end{tabular}   \\ \hline
                & Level 5             & \begin{tabular}[c]{@{}l@{}}Conv 9\\ Conv 10\end{tabular}  & \begin{tabular}[c]{@{}l@{}}3 $\times\:$3/256\\ 3 $\times\:$3/256\end{tabular} & \begin{tabular}[c]{@{}c@{}}1\\ 1\end{tabular} & \begin{tabular}[c]{@{}c@{}}32 $\times\:$16 $\times\:$256\\ 32 $\times\:$16 $\times\:$256\end{tabular} \\ \hline
                & Level 6             & \begin{tabular}[c]{@{}l@{}}Conv 11\\ Conv 12\end{tabular} & \begin{tabular}[c]{@{}l@{}}3 $\times\:$3/128\\ 3 $\times\:$3/128\end{tabular} & \begin{tabular}[c]{@{}c@{}}1\\ 1\end{tabular} & \begin{tabular}[c]{@{}c@{}}64 $\times\:$32 $\times\:$128\\ 64 $\times\:$32 $\times\:$128\end{tabular} \\ \hline
                & Level 7             & \begin{tabular}[c]{@{}l@{}}Conv 13\\ Conv 14\end{tabular} & \begin{tabular}[c]{@{}l@{}}3 $\times\:$3/64\\ 3 $\times\:$3/64\end{tabular}   & \begin{tabular}[c]{@{}c@{}}1\\ 1\end{tabular} & \begin{tabular}[c]{@{}c@{}}128 $\times\:$64 $\times\:$64\\ 128 $\times\:$64 $\times\:$64\end{tabular} \\ \hline
\textbf{Output} &                     & Conv 15                                                   & 1 $\times\:$1                                                         & 1                                             & 128 $\times\:$64 $\times\:$1                                                          \\ \hline
\end{tabular}
}
\label{unet_detail_1}
\end{table}

\subsubsection{Reconstruction of the nodule shape}

After obtaining the nodule segmentation for each plane , we apply the consensus module to calculate the final segmentation. The value of $k_{th}$ consensus pixel $c_k$ of segmentation mask is calculated as

\begin{equation}
c_k = \Gamma\left[ \sum_{i=0}^{M-1} S_{k_i} \right], where\:k\in[0,N-1] 
\end{equation}
\newline where $S$ represents the estimated segmentation, $M$ is the number of estimations (three in our case, axial, coronal, and sagittal), $N$ is the number of voxels in the VOI, and $\Gamma$ is defined as

\begin{equation} \Gamma(g) = \begin{cases} 
      1, & if\: g \geq (M*C_R) \\
      0, & Otherwise 
   \end{cases}
\end{equation}
\newline
Here, $C_R$ is the consensus ratio set as 0.5 to represent a 50\% consensus.

\section{Experimental Setup}
\label{ES}
\subsection{Dataset and Pre-Processing}
\label{pre_processing}

All the experiments were carried out in accordance with relevant guidelines. We acknowledge the National Cancer Institute and the Foundation for the National Institutes of Health, and their critical role in the creation of the free publicly available database; the Lung Image Database Consortium and Image Database Resource Initiative (LIDC-IDRI) Database \cite{mcnitt2007lung,armato2011lung} used in this study is freely available to browse, download, and use for commercial, scientific and educational purposes as outlined in the Creative Commons Attribution 3.0 Unported License.

The LIDC-IDRI is the largest repository of CT scans to facilitate computer-aided systems on the assessment of lung nodule detection, classification and quantification. This dataset is composed of 1,018 cases of diagnostic and lung cancer screening thoracic CT scans with marked-up annotated lesions belonging to 1,010 patients. Each subject in the dataset includes images from a clinical thoracic CT scan and the results of a two-phase image annotation process performed by four experienced thoracic radiologists. We consider the nodules that are annotated by all the four radiologists and have a diameter of no less than 3mm, as in previous studies \cite{feng2017discriminative}, \cite{wu2018joint}. Because of the inter-variability among the four radiologists, a 50\% consensus criterion \cite{kubota2011segmentation} is opted to generate the ground truth boundary of the pulmonary nodule segmentation and a python library named pyLIDC is used for this purpose. A total of 893 nodules are selected and randomly divided into three subsets, including training, validation, and testing sets  contain 356, 45, and 492 nodules, respectively. 

Unlike previous works \cite{wang2017central,wang2017multi,tong2018improved,amorim2019lung} that, used a fixed margin strategy while extracting the ROIs for training purpose, we used random margins strategy. As discussed above, in the fixed margin approach, the nodule appears at the center of the ROI, which constrains the model learning for the possibility of nodules existing at the corners of an ROI. Because our focus is to detect the nodule presence anywhere within a given ROI, we prepared our data following a random margin strategy. The values of all margins shown in Figure \ref{DROI_mask} are generated with a random function and restricted between zero and $f$ defined as

\begin{equation}
    f = D_{max} * R_T
\end{equation}
\newline
where $D_{max}$ is the maximum diameter of a nodule within the slice. For patch-wise investigations of a nodule, the network should learn about the absence of nodules in a given ROI, so we also include the non-nodule-containing ROI from both sides (i.e., after the top and bottom slices) of the nodule ROI.

\subsection{Implementation details}

The proposed model was implemented using Keras \cite{chollet2015keras} framework and optimized by minimizing Equation \ref{eq_2} through SGD algorithm, and trained on 12,821 images sized $128\times128$. We begin training the model with random weights and with learning rate of $10^{-4}$. We use mini-batch size of 8 on a NVIDIA TESLA V100 TENSOR CORE GPU. The networks converges within $700$ epochs.

\subsection{Evaluation parameters}
\label{ep}
We use following evaluation parameters to evaluate the perform of our proposed method.

\subsubsection*{Dice Similarity Coefficient}
Equation \ref{eq_3} define the dice similarity coefficient (DSC), which is widely used parameter to evaluate the degree of overlap of predicted segment ($S_{Pred}$) with reference segment ($S_{Ref}$) \cite{jung2018ground, wang2017central}. The DSC values ranges [0,1], while 0 and 1 indicate no overlap and complete overlap, respectively. 

\begin{equation}
    DSC = \frac{2*S_{Pred}\cap S_{Ref}}{S_{Pred} + S_{Ref}}
    \label{eq_3}
\end{equation}

\subsubsection{Sensitivity and Positive Predictive Value}
The pixel classification performance and correctness of the segmentation area are measured by the sensitivity (SEN) and the positive predictive value (PPV), which are defined as 

\begin{equation}
    SEN = \frac{S_{Pred}\cap S_{Ref}}{S_{Ref}}
\end{equation}

\begin{equation}
    PPV = \frac{S_{Pred}\cap S_{Ref}}{S_{Pred}}
\end{equation}

\section{Results and Discussion}

\begin{table*}[b]
\centering
\caption{The mean $\pm$ standard deviation for quantitative results of various segmentation methods with the best performance indicated in bold.}
\scalebox{1.2}{
\begin{tabular}{llll}
\hline
\multicolumn{1}{c}{\textbf{Methodology}}                                                               & \textbf{DSC (\%)} & \textbf{SEN (\%)} & \textbf{PPV (\%)} \\ \hline
Central focused CNN \cite{wang2017central}                                                                             & 78.55 $\pm$ 12.49     & 86.01 $\pm$ 15.22     & 75.79 $\pm$ 14.73     \\ \hline
Multi-crop CNN  \cite{shen2017multi}                                                                                 & 77.51 $\pm$ 11.40     & 88.83 $\pm$ 12.34     & 71.42 $\pm$ 14.78     \\ \hline
Multi-view CNN \cite{kang20173d}                                                                                  & 75.89 $\pm$ 12.99     & 87.16 $\pm$ 12.91     & 70.81 $\pm$ 17.57     \\ \hline
Multi-view deep CNN \cite{wang2017multi}                                                                             & 77.85 $\pm$ 12.94     & 86.96 $\pm$ 15.73     & 77.33 $\pm$ 13.26     \\ \hline
\begin{tabular}[c]{@{}l@{}}Multichannel ROI Based on \\ Deep Structured Algorithms \cite{sun2017automatic}\end{tabular} & 77.01 $\pm$ 12.93     & 85.45 $\pm$ 15.97     & 73.52 $\pm$ 14.62     \\ \hline
Cascaded dual-pathway Res-Net \cite{liu2019cascaded}                                                                   & 81.58 $\pm$ 11.05     & 87.30 $\pm$ 14.30     & 79.71 $\pm$ 13.59     \\ \hline
Unsupervised metaheuristic search \cite{shakibapour2019unsupervised}                                                               & 82.34 $\pm$ 5.40      & 87.10 $\pm$ 9.78      & 85.59 $\pm$ 11.06     \\ \hline
Constant ROI with Multi-View Deep Residual Learning                                                                                  & 84.35 $\pm$ 11.72        & 89.02 $\pm$ 8.91       & 86.73 $\pm$ 10.11       \\ \hline
\textbf{A-ROI with Multi-View Deep Residual Learning}                                                                                  & \textbf{87.55 $\pm$ 10.58   }     & \textbf{91.62 $\pm$ 8.47   }     & \textbf{88.24 $\pm$ 9.52   }     \\ \hline
\end{tabular}
}
\label{overall_t}
\end{table*}

\label{RD}
We performed various experiments on the LIDC-IDRI dataset to evaluate the performance of our approach . The results were analyzed based on aspects of the effect of the $R_T$  value, overall performance, robustness, and visual performance. The following subsections review our analysis.

\subsection{Performance Sensitivity to the $R_T$ value}
\begin{figure}[t]
\centering
%captionsetup{justification=centering}
\centerline{\includegraphics[width=.5\textwidth]{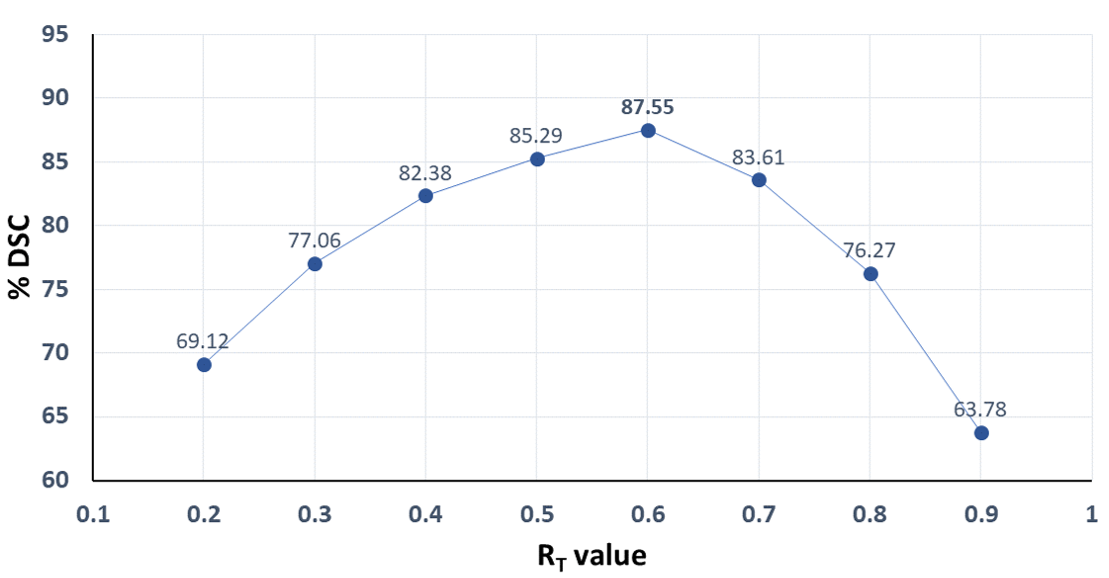}}
\caption{A graph of various values of $R_T$ and the corresponding performance in terms of \% DSC on the LIDC-IDRI testing set.}
\label{rt_graph}
\end{figure}

The value of $R_T$ is determined experimentally, so multiple experiments were performed with various values of $R_T$ to obtain its optimum value. Figure \ref{rt_graph} shows a range of $R_T$ values and their respective overall performance. With an increase in the value of $R_T$, the performance also increases until it reaches a specific point (i.e., 0.6), after which the performance decreases. The graph illustrates that for lower $R_T$ representing a larger ROI, the performance is degraded because the ROI is not focused on a single nodule and includes redundant information or non-nodule structures. This extraneous information makes the segmentation process challenging, consequently reducing overall performance. On the other hand, when a high value of $R_T$ is selected, the ROI not only focuses on a nodule but also remains very close to the nodule boundary, which is insufficient to handle small displacements of the nodule across adjacent slices. Therefore, a moderate value of $R_T$ provides the best performance, and we used the same value of 0.6 for $R_T$ in all experiments reported.

\subsection{Overall Performance}
The overall performance of the proposed method is evaluated by using all evaluation parameters mentioned above. Table \ref{overall_t} lists these results for various methods, and the adaptive ROI method significantly outperforms previous state-of-the-art techniques. To demonstrate the effectiveness of adaptive ROI, we also applied a constant ROI with Multi-view Residual leaning, which also outperformed the existing techniques reflecting the power of residual learning, as it has not been previously used for nodule segmentation.

The results in Table 3 also suggest that the incorporation of adaptive ROI significantly improves the results, with the average DCS value increasing by approximately 3\%. The reason for such high performance is due to the two advantages, as explained in Figure \ref{ROIs}. First, the approach reduces the size of the ROI enabling it to focus only on the nodule and eliminating possible redundant information (i.e., similar non-nodule structures) from the ROI, and, second, it assists the residual network in the classification of the nodule and non-nodule voxels. On the other hand, the constant ROI must cover a large area (i.e., surrounding of nodule) to incorporate the entire shape of the nodule, which impacts the performance of the network.

\begin{figure}[t]
\centering
%captionsetup{justification=centering}
\centerline{\includegraphics[width=.5\textwidth]{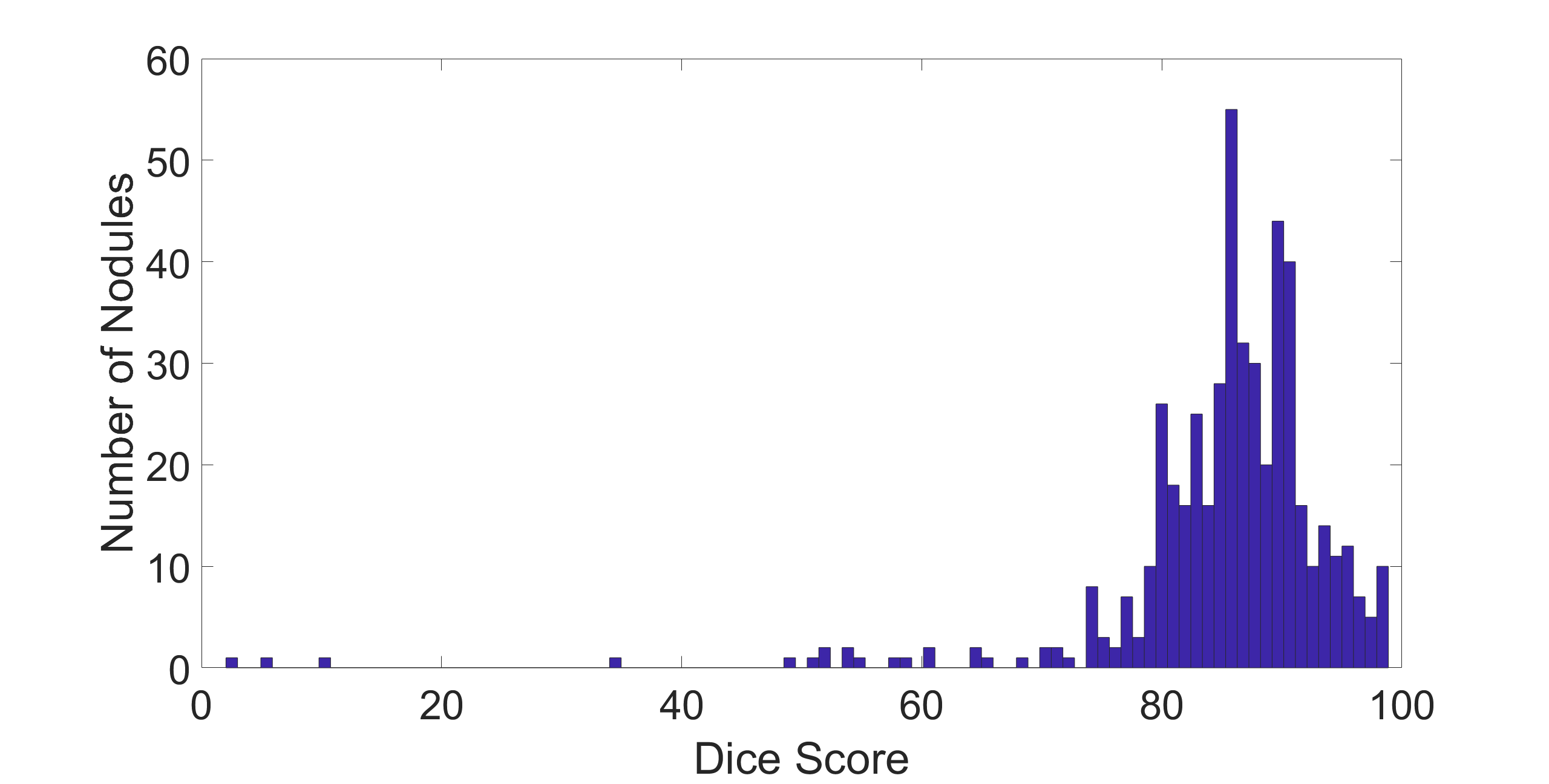}}
\caption{Distribution of DSC for the LIDC-IDRI test set.}
\label{hist_result}
\end{figure}

Figure \ref{hist_result} demonstrates the distribution of dice scores achieved on the test set of the LIDC-IDRI dataset showing that most test instances are above 85\%, indicating the excellent performance of our proposed method.

\subsection{Evaluation of Robustness}

The LIDC-IDRI dataset also provides annotations about the levels of various characteristics of the nodules, such as subtlety, internal structure, calcification, sphericity, margin, lobulation, speculation, texture, and malignancy. These characteristics represent the levels of difficulty for detecting the nodule boundary. To evaluate the robustness of our method, we partitioned the test data into a different level of each characteristic and extracted the results against each level, as presented in Table \ref{robust_t}. The dice score achieved for each group is similar, indicating the robustness of our approach.

\begin{table*}[h]
\centering
\caption{Average DSC for different nodule types in the LIDC-IDRI testing set.}
\scalebox{1.2}{
\begin{tabular}{|l|l|l|l|l|l|c|}
\hline
\multicolumn{1}{|c|}{\multirow{2}{*}{\textbf{Characteristics}}} & \multicolumn{6}{l|}{\textbf{Characteristic scores}}                                                                                               \\ \cline{2-7} 
\multicolumn{1}{|c|}{}                                          & 1                       & 2                       & 3                       & 4                & 5                       & \multicolumn{1}{l|}{6} \\ \hline
\textbf{Calcification}                                          & \multicolumn{1}{c|}{\_} & \multicolumn{1}{c|}{\_} & 84.61 {[}18{]}          & 83.88  {[}42{]}  & 86.24  {[}27{]}         & 88.15  {[}405{]}       \\ \hline
\textbf{Internal structure}                                     & 87.62  {[}487{]}        & 79.27  {[}3{]}          & \multicolumn{1}{c|}{\_} & 81.48  {[}2{]}   & \multicolumn{1}{c|}{\_} & \_                     \\ \hline
\textbf{Lobulation}                                             & 86.96  {[}201{]}        & 88.78  {[}164{]}        & 86.56  {[}78{]}         & 85.94  {[}31{]}  & 89.74  {[}18{]}         & \_                     \\ \hline
\textbf{Malignancy}                                             & 85.67  {[}39{]}         & 86.56  {[}114{]}        & 89.26  {[}163{]}        & 86.67  {[}98{]}  & 87.52  {[}78{]}         & \_                     \\ \hline
\textbf{Margin}                                                 & 86.42  {[}9{]}          & 85.61  {[}37{]}         & 86.73  {[}78{]}         & 88.21  {[}232{]} & 87.51  {[}136{]}        & \_                     \\ \hline
\textbf{Sphericity}                                             & \multicolumn{1}{c|}{\_} & 86.79  {[}38{]}         & 85.28  {[}153{]}        & 89.16  {[}218{]} & 88.12  {[}83{]}         & \_                     \\ \hline
\textbf{Speculation}                                            & 89.29  {[}257{]}        & 85.18  {[}165{]}        & 86.96  {[}32{]}         & 86.37  {[}14{]}  & 86.68  {[}24{]}         & \_                     \\ \hline
\textbf{Subtlety}                                               & 80.52  {[}4{]}          & 82.65  {[}22{]}         & 87.51  {[}131{]}        & 87.09  {[}238{]} & 90.17  {[}97{]}         & \_                     \\ \hline
\textbf{Texture}                                                & 82.18  {[}11{]}         & 85.67  {[}18{]}         & 86.54  {[}26{]}         & 87.55  {[}107{]} & 87.93  {[}330{]}        & \_                     \\ \hline
\end{tabular}
}
\label{robust_t}
\end{table*}

\begin{figure*}[!b]
\centering
%captionsetup{justification=centering}
\centerline{\includegraphics[width=.9\textwidth]{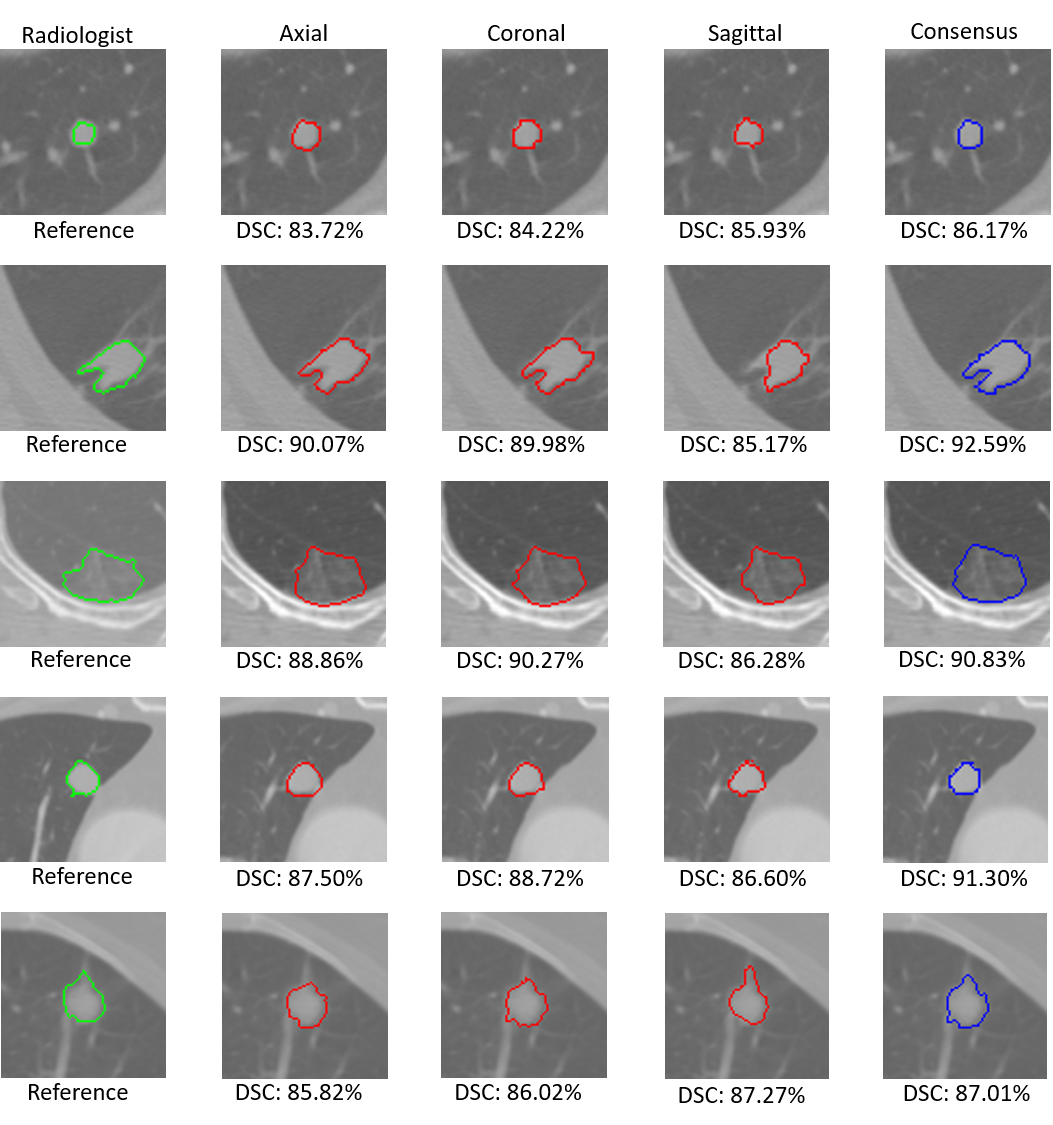}}
\caption{The segmentation results of proposed approach, at different stages (i.e., after employing segmentation along axial, coronal, sagittal axis) and finally the segmentation outputs of consensus module have been presented with their respective dice scores.}
\label{v_result}
\end{figure*}

\begin{table}[h]
\centering
\caption{Average DSC for each stage of the proposed approach.}
\scalebox{1}{
\begin{tabular}{lllll}
\hline
                    & \textbf{Axial}            & \textbf{Coronal}          & \textbf{Sagittal}         & \textbf{Consensus}        \\ \hline
\textbf{Dice Score} & \multicolumn{1}{c}{85.29} & \multicolumn{1}{c}{84.76} & \multicolumn{1}{c}{83.58} & \multicolumn{1}{c}{87.55} \\ \hline
\end{tabular}
}
\label{comp_overall}
\end{table}

\begin{figure*}[t]
\centering
%captionsetup{justification=centering}
\centerline{\includegraphics[width=1.0\textwidth]{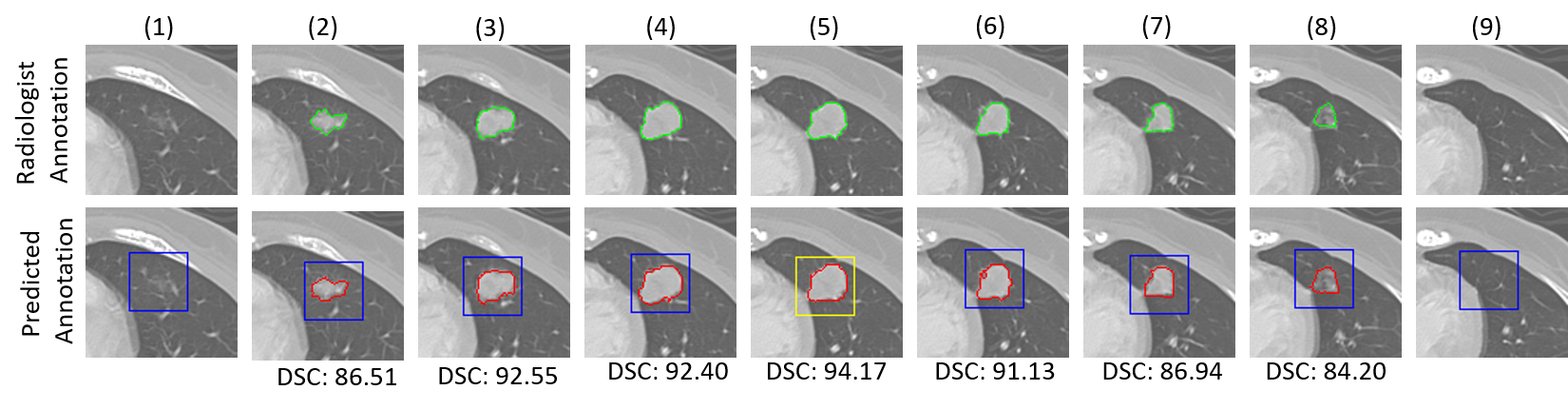}}
\caption{The segmentation performance of our proposed approach on each slice of a single nodule with the corresponding dice scores. From columns 1-9, the sequence of slices is observed with the nodule seen in columns 2-7 and no nodule in columns 1 and 9. The squares marked in the predicted segmentation row depict the ROIs, where yellow represents the initial (manually provided) ROI, and blue represents the adaptive ROIs.}
\label{single_nodule}
\end{figure*}

\subsection{Visual analysis}
The results of our method are visualized to analyze its performance at various stages of the proposed approach. In Figure \ref{v_result}, the segmentation outputs, following the axial, coronal, and sagittal view investigations, are shown with the output of the consensus module presented in the last column. Nodules of various types from test data are randomly selected, including the attached, non-attached, and nodule with extreme subtlety. The results suggest that adaptive ROI with residual learning enhances the performance of the axial view investigation. However, in the cases where the penetration of the nodule along the z-axis is higher than the presence of the nodule on the x-y plane, the performance along the sagittal and coronal axes is only slightly higher. Subsequently, in a few cases, the performance following the consensus module is also slightly reduced. However, according to the dice scores of the outputs following the axial, coronal, sagittal, and consensus modules listed in Table \ref{comp_overall}, the performance still improves significantly after applying the consensus module.

Because our method automatically extracts the VOI, an estimation of nodule penetration in the surrounding slices must be evaluated. Figure \ref{single_nodule} shows the visual results from a complete nodule. As described above, a manual ROI is needed to initialize the segmentation process, which is presented in yellow, and the adaptive ROIs are calculated through the A-ROI algorithm for the surrounding slices, as presented in blue. The detected presence of a nodule in each slice is highlighted with the corresponding dice scores. Our method successfully detects the penetration of the nodule in each slice and stops the investigation immediately after the end of the nodule is reached from both sides.

\section*{Conclusion}
\label{co}
We introduced a novel, two-stage pulmonary nodule segmentation technique that generates a highly accurate, 3-D segmentation of a nodule with minimum human interaction. First, the technique requires a 2-D ROI along the axial axis containing the nodule as input and then automatically estimates the volume of interest (VOI). For VOI extraction, we proposed the novel adaptive ROI algorithm with a Deep Residual U-Net architecture to leverage the position of the nodule within the ROI of the current slice to adjust the position and shape of the ROI for the next slice. Second, additional patch-wise segmentations of the nodule along the coronal and sagittal axes are performed by applying two Residual U-Nets. Finally, the segmentation outputs of the axial, coronal, and sagittal axes are processed through the consensus module to generate the final segmentation mask. Our technique was evaluated on the LIDC-IDRI dataset with quantitative and visual results. We demonstrated that our approach outperforms the previous state-of-the-art techniques in terms of segmentation dice scores and is significantly robust for various nodule types, suggesting its suitability for real-time implementation to minimize the radiologist effort through the automatic estimation of the VOI and to enhance the accuracy of segmentation. Future research will include an extension of the framework to perform a time-based analysis of a selected nodule by using sequential scans of a patient, which will assist radiologists in estimating the malignancy of an identified nodule by observing its development over time.

\bibliographystyle{IEEEtran}
% \bibliography{sample.bib}

\begin{thebibliography}{10}
\providecommand{\url}[1]{#1}
\csname url@samestyle\endcsname
\providecommand{\newblock}{\relax}
\providecommand{\bibinfo}[2]{#2}
\providecommand{\BIBentrySTDinterwordspacing}{\spaceskip=0pt\relax}
\providecommand{\BIBentryALTinterwordstretchfactor}{4}
\providecommand{\BIBentryALTinterwordspacing}{\spaceskip=\fontdimen2\font plus
\BIBentryALTinterwordstretchfactor\fontdimen3\font minus
  \fontdimen4\font\relax}
\providecommand{\BIBforeignlanguage}[2]{{%
\expandafter\ifx\csname l@#1\endcsname\relax
\typeout{** WARNING: IEEEtran.bst: No hyphenation pattern has been}%
\typeout{** loaded for the language `#1'. Using the pattern for}%
\typeout{** the default language instead.}%
\else
\language=\csname l@#1\endcsname
\fi
#2}}
\providecommand{\BIBdecl}{\relax}
\BIBdecl

\bibitem{torre2015global}
L.~A. Torre, F.~Bray, R.~L. Siegel, J.~Ferlay, J.~Lortet-Tieulent, and
  A.~Jemal, ``Global cancer statistics, 2012,'' \emph{CA: a cancer journal for
  clinicians}, vol.~65, no.~2, pp. 87--108, 2015.

\bibitem{malvezzi2019european}
M.~Malvezzi, G.~Carioli, P.~Bertuccio, P.~Boffetta, F.~Levi, C.~La~Vecchia, and
  E.~Negri, ``European cancer mortality predictions for the year 2019 with
  focus on breast cancer,'' \emph{Annals of Oncology}, vol.~30, no.~5, pp.
  781--787, 2019.

\bibitem{blandin2017progress}
S.~Blandin~Knight, P.~A. Crosbie, H.~Balata, J.~Chudziak, T.~Hussell, and
  C.~Dive, ``Progress and prospects of early detection in lung cancer,''
  \emph{Open biology}, vol.~7, no.~9, p. 170070, 2017.

\bibitem{mozley2012measurement}
P.~D. Mozley, C.~Bendtsen, B.~Zhao, L.~H. Schwartz, M.~Thorn, Y.~Rong,
  L.~Zhang, A.~Perrone, R.~Korn, and A.~J. Buckler, ``Measurement of tumor
  volumes improves recist-based response assessments in advanced lung cancer,''
  \emph{Translational oncology}, vol.~5, no.~1, p.~19, 2012.

\bibitem{devaraj2017use}
A.~Devaraj, B.~van Ginneken, A.~Nair, and D.~Baldwin, ``Use of volumetry for
  lung nodule management: theory and practice,'' \emph{Radiology}, vol. 284,
  no.~3, pp. 630--644, 2017.

\bibitem{moltz2009advanced}
J.~H. Moltz, L.~Bornemann, J.-M. Kuhnigk, V.~Dicken, E.~Peitgen, S.~Meier,
  H.~Bolte, M.~Fabel, H.-C. Bauknecht, M.~Hittinger \emph{et~al.}, ``Advanced
  segmentation techniques for lung nodules, liver metastases, and enlarged
  lymph nodes in ct scans,'' \emph{IEEE Journal of selected topics in signal
  processing}, vol.~3, no.~1, pp. 122--134, 2009.

\bibitem{wu2019survey}
J.~Wu and T.~Qian, ``A survey of pulmonary nodule detection, segmentation and
  classification in computed tomography with deep learning techniques,''
  \emph{Journal of Medical Artificial Intelligence}, vol.~2, 2019.

\bibitem{wang2017central}
S.~Wang, M.~Zhou, Z.~Liu, Z.~Liu, D.~Gu, Y.~Zang, D.~Dong, O.~Gevaert, and
  J.~Tian, ``Central focused convolutional neural networks: Developing a
  data-driven model for lung nodule segmentation,'' \emph{Medical image
  analysis}, vol.~40, pp. 172--183, 2017.

\bibitem{shakibapour2019unsupervised}
E.~Shakibapour, A.~Cunha, G.~Aresta, A.~M. Mendon{\c{c}}a, and A.~Campilho,
  ``An unsupervised metaheuristic search approach for segmentation and volume
  measurement of pulmonary nodules in lung ct scans,'' \emph{Expert Systems
  with Applications}, vol. 119, pp. 415--428, 2019.

\bibitem{liu2019cascaded}
H.~Liu, H.~Cao, E.~Song, G.~Ma, X.~Xu, R.~Jin, Y.~Jin, and C.-C. Hung, ``A
  cascaded dual-pathway residual network for lung nodule segmentation in ct
  images,'' \emph{Physica Medica}, vol.~63, pp. 112--121, 2019.

\bibitem{armato2011lung}
S.~G. Armato~III, G.~McLennan, L.~Bidaut, M.~F. McNitt-Gray, C.~R. Meyer, A.~P.
  Reeves, B.~Zhao, D.~R. Aberle, C.~I. Henschke, E.~A. Hoffman \emph{et~al.},
  ``The lung image database consortium (lidc) and image database resource
  initiative (idri): a completed reference database of lung nodules on ct
  scans,'' \emph{Medical physics}, vol.~38, no.~2, pp. 915--931, 2011.

\bibitem{dehmeshki2008segmentation}
J.~Dehmeshki, H.~Amin, M.~Valdivieso, and X.~Ye, ``Segmentation of pulmonary
  nodules in thoracic ct scans: a region growing approach,'' \emph{IEEE
  transactions on medical imaging}, vol.~27, no.~4, pp. 467--480, 2008.

\bibitem{diciotti20083}
S.~Diciotti, G.~Picozzi, M.~Falchini, M.~Mascalchi, N.~Villari, and G.~Valli,
  ``3-d segmentation algorithm of small lung nodules in spiral ct images,''
  \emph{IEEE transactions on Information Technology in Biomedicine}, vol.~12,
  no.~1, pp. 7--19, 2008.

\bibitem{diciotti2011automated}
S.~Diciotti, S.~Lombardo, M.~Falchini, G.~Picozzi, and M.~Mascalchi,
  ``Automated segmentation refinement of small lung nodules in ct scans by
  local shape analysis,'' \emph{IEEE Transactions on Biomedical Engineering},
  vol.~58, no.~12, pp. 3418--3428, 2011.

\bibitem{rendon2016automatic}
E.~Rendon-Gonzalez and V.~Ponomaryov, ``Automatic lung nodule segmentation and
  classification in ct images based on svm,'' in \emph{2016 9th International
  Kharkiv Symposium on Physics and Engineering of Microwaves, Millimeter and
  Submillimeter Waves (MSMW)}.\hskip 1em plus 0.5em minus 0.4em\relax IEEE,
  2016, pp. 1--4.

\bibitem{wang2017adaptive}
D.~Wang, J.~Wang, Y.~Du, and P.~Tang, ``Adaptive solitary pulmonary nodule
  segmentation for digital radiography images based on random walks and
  sequential filter,'' \emph{IEEE Access}, vol.~5, pp. 1460--1468, 2017.

\bibitem{jung2018ground}
J.~Jung, H.~Hong, and J.~M. Goo, ``Ground-glass nodule segmentation in chest ct
  images using asymmetric multi-phase deformable model and pulmonary vessel
  removal,'' \emph{Computers in biology and medicine}, vol.~92, pp. 128--138,
  2018.

\bibitem{shakir20183}
H.~Shakir, T.~M.~R. Khan, and H.~Rasheed, ``3-d segmentation of lung nodules
  using hybrid level sets,'' \emph{Computers in biology and medicine}, vol.~96,
  pp. 214--226, 2018.

\bibitem{latif2018phonocardiographic}
S.~Latif, M.~Usman, R.~Rana, and J.~Qadir, ``Phonocardiographic sensing using
  deep learning for abnormal heartbeat detection,'' \emph{IEEE Sensors
  Journal}, vol.~18, no.~22, pp. 9393--9400, 2018.

\bibitem{latif2018mobile}
S.~Latif, M.~Y. Khan, A.~Qayyum, J.~Qadir, M.~Usman, S.~M. Ali, Q.~H. Abbasi,
  and M.~A. Imran, ``Mobile technologies for managing non-communicable diseases
  in developing countries,'' in \emph{Mobile Applications and Solutions for
  Social Inclusion}.\hskip 1em plus 0.5em minus 0.4em\relax IGI Global, 2018,
  pp. 261--287.

\bibitem{usman2017using}
M.~Usman, S.~Latif, and J.~Qadir, ``Using deep autoencoders for facial
  expression recognition,'' in \emph{2017 13th International Conference on
  Emerging Technologies (ICET)}.\hskip 1em plus 0.5em minus 0.4em\relax IEEE,
  2017, pp. 1--6.

\bibitem{latif2018cross}
S.~Latif, A.~Qayyum, M.~Usman, and J.~Qadir, ``Cross lingual speech emotion
  recognition: Urdu vs. western languages,'' in \emph{2018 International
  Conference on Frontiers of Information Technology (FIT)}.\hskip 1em plus
  0.5em minus 0.4em\relax IEEE, 2018, pp. 88--93.

\bibitem{latif2018automating}
S.~Latif, M.~Asim, M.~Usman, J.~Qadir, and R.~Rana, ``Automating motion
  correction in multishot mri using generative adversarial networks,''
  \emph{arXiv preprint arXiv:1811.09750}, 2018.

\bibitem{usman2019motion}
M.~Usman, S.~Latif, M.~Asim, and J.~Qadir, ``Motion corrected multishot mri
  reconstruction using generative networks with sensitivity encoding,''
  \emph{arXiv preprint arXiv:1902.07430}, 2019.

\bibitem{litjens2017survey}
G.~Litjens, T.~Kooi, B.~E. Bejnordi, A.~A.~A. Setio, F.~Ciompi, M.~Ghafoorian,
  J.~A. Van Der~Laak, B.~Van~Ginneken, and C.~I. S{\'a}nchez, ``A survey on
  deep learning in medical image analysis,'' \emph{Medical image analysis},
  vol.~42, pp. 60--88, 2017.

\bibitem{guo2018review}
Y.~Guo, Y.~Liu, T.~Georgiou, and M.~S. Lew, ``A review of semantic segmentation
  using deep neural networks,'' \emph{International journal of multimedia
  information retrieval}, vol.~7, no.~2, pp. 87--93, 2018.

\bibitem{usman2019volumetric}
M.~Usman, B.-D. Lee, S.~S. Byon, S.~H. Kim, and B.~IlLee, ``Volumetric lung
  nodule segmentation using adaptive roi with multi-view residual learning,''
  \emph{arXiv preprint arXiv:1912.13335}, 2019.

\bibitem{rocha2019comparison}
J.~Rocha, A.~Cunha, and A.~M. Mendon{\c{c}}a, ``Comparison of conventional and
  deep learning based methods for pulmonary nodule segmentation in ct images,''
  in \emph{EPIA Conference on Artificial Intelligence}.\hskip 1em plus 0.5em
  minus 0.4em\relax Springer, 2019, pp. 361--371.

\bibitem{ronneberger2015u}
O.~Ronneberger, P.~Fischer, and T.~Brox, ``U-net: Convolutional networks for
  biomedical image segmentation,'' in \emph{International Conference on Medical
  image computing and computer-assisted intervention}.\hskip 1em plus 0.5em
  minus 0.4em\relax Springer, 2015, pp. 234--241.

\bibitem{wang2017multi}
S.~Wang, M.~Zhou, O.~Gevaert, Z.~Tang, D.~Dong, Z.~Liu, and J.~Tian, ``A
  multi-view deep convolutional neural networks for lung nodule segmentation,''
  in \emph{2017 39th Annual International Conference of the IEEE Engineering in
  Medicine and Biology Society (EMBC)}.\hskip 1em plus 0.5em minus 0.4em\relax
  IEEE, 2017, pp. 1752--1755.

\bibitem{tong2018improved}
G.~Tong, Y.~Li, H.~Chen, Q.~Zhang, and H.~Jiang, ``Improved u-net network for
  pulmonary nodules segmentation,'' \emph{Optik}, vol. 174, pp. 460--469, 2018.

\bibitem{amorim2019lung}
P.~H. Amorim, T.~F. de~Moraes, J.~V. da~Silva, and H.~Pedrini, ``Lung nodule
  segmentation based on convolutional neural networks using multi-orientation
  and patchwise mechanisms,'' in \emph{ECCOMAS Thematic Conference on
  Computational Vision and Medical Image Processing}.\hskip 1em plus 0.5em
  minus 0.4em\relax Springer, 2019, pp. 286--295.

\bibitem{hancock2019lung}
M.~C. Hancock and J.~F. Magnan, ``Lung nodule segmentation via level set
  machine learning,'' \emph{arXiv preprint arXiv:1910.03191}, 2019.

\bibitem{zhang2018road}
Z.~Zhang, Q.~Liu, and Y.~Wang, ``Road extraction by deep residual u-net,''
  \emph{IEEE Geoscience and Remote Sensing Letters}, vol.~15, no.~5, pp.
  749--753, 2018.

\bibitem{zou2004statistical}
K.~H. Zou, S.~K. Warfield, A.~Bharatha, C.~M. Tempany, M.~R. Kaus, S.~J. Haker,
  W.~M. Wells~III, F.~A. Jolesz, and R.~Kikinis, ``Statistical validation of
  image segmentation quality based on a spatial overlap index1: scientific
  reports,'' \emph{Academic radiology}, vol.~11, no.~2, pp. 178--189, 2004.

\bibitem{mcnitt2007lung}
M.~F. McNitt-Gray, S.~G. Armato~III, C.~R. Meyer, A.~P. Reeves, G.~McLennan,
  R.~C. Pais, J.~Freymann, M.~S. Brown, R.~M. Engelmann, P.~H. Bland
  \emph{et~al.}, ``The lung image database consortium (lidc) data collection
  process for nodule detection and annotation,'' \emph{Academic radiology},
  vol.~14, no.~12, pp. 1464--1474, 2007.

\bibitem{feng2017discriminative}
X.~Feng, J.~Yang, A.~F. Laine, and E.~D. Angelini, ``Discriminative
  localization in cnns for weakly-supervised segmentation of pulmonary
  nodules,'' in \emph{International Conference on Medical Image Computing and
  Computer-Assisted Intervention}.\hskip 1em plus 0.5em minus 0.4em\relax
  Springer, 2017, pp. 568--576.

\bibitem{wu2018joint}
B.~Wu, Z.~Zhou, J.~Wang, and Y.~Wang, ``Joint learning for pulmonary nodule
  segmentation, attributes and malignancy prediction,'' in \emph{2018 IEEE 15th
  International Symposium on Biomedical Imaging (ISBI 2018)}.\hskip 1em plus
  0.5em minus 0.4em\relax IEEE, 2018, pp. 1109--1113.

\bibitem{kubota2011segmentation}
T.~Kubota, A.~K. Jerebko, M.~Dewan, M.~Salganicoff, and A.~Krishnan,
  ``Segmentation of pulmonary nodules of various densities with morphological
  approaches and convexity models,'' \emph{Medical Image Analysis}, vol.~15,
  no.~1, pp. 133--154, 2011.

\bibitem{chollet2015keras}
F.~Chollet \emph{et~al.}, ``Keras,'' \url{https://github.com/fchollet/keras},
  2015.

\bibitem{shen2017multi}
W.~Shen, M.~Zhou, F.~Yang, D.~Yu, D.~Dong, C.~Yang, Y.~Zang, and J.~Tian,
  ``Multi-crop convolutional neural networks for lung nodule malignancy
  suspiciousness classification,'' \emph{Pattern Recognition}, vol.~61, pp.
  663--673, 2017.

\bibitem{kang20173d}
G.~Kang, K.~Liu, B.~Hou, and N.~Zhang, ``3d multi-view convolutional neural
  networks for lung nodule classification,'' \emph{PloS one}, vol.~12, no.~11,
  p. e0188290, 2017.

\bibitem{sun2017automatic}
W.~Sun, B.~Zheng, and W.~Qian, ``Automatic feature learning using multichannel
  roi based on deep structured algorithms for computerized lung cancer
  diagnosis,'' \emph{Computers in biology and medicine}, vol.~89, pp. 530--539,
  2017.
  
\end{thebibliography}

\end{document}